\documentclass[sn-basic]{sn-jnl}
\usepackage{graphicx}%
\usepackage{multirow}%
\usepackage{amsmath,amssymb,amsfonts}%
\usepackage{amsthm}%
\usepackage{mathrsfs}%
\usepackage[title]{appendix}%
\usepackage{xcolor}%
\usepackage{textcomp}%
\usepackage{manyfoot}%
\usepackage{booktabs}%
\usepackage{algorithm}%
\usepackage{algorithmicx}%
\usepackage{algpseudocode}%
\usepackage{listings}%
\usepackage{bm}
\usepackage{subcaption}
\usepackage{hyperref}

\begin{document}
\title[Article Title]{Formation of a Bose Star in a Rotating Cloud}

\author*[1,2]{\sur{Kuldeep J. Purohit }}\email{kuldeepjrajpurohit@gmail.com}

\author[3,4,5,6]{\sur{Pravin Kumar Natwariya}}\email{pvn.sps@gmail.com}
\equalcont{These authors contributed equally to this work.}

\author[3]{\sur{Jitesh R. Bhatt}}\email{jeet@prl.res.in}
\equalcont{These authors contributed equally to this work.}

\author[1]{\sur{Prashant K. Mehta}}\email{pk.mehta-phy@msubaroda.co.in}
\equalcont{These authors contributed equally to this work.}

\affil*[1]{\orgdiv{Physics Department}, \orgname{Faculty of Science,
    The Maharaja Sayajirao University of Baroda},  \city{Vadodara}, \postcode{390 002}, \state{Gujarat}, \country{India}}

\affil[2]{\orgdiv{Department of Physics}, \orgname{St. Xavier's College},  \city{Ahmedabad}, \postcode{380 009}, \state{Gujarat}, \country{India}}

\affil[3]{\orgdiv{Theoretical Physics Division}, \orgname{Physical Research Laboratory}, \city{Ahmedabad}, \postcode{380 009}, \state{Gujarat}, \country{India}}

\affil[4]{\orgdiv{Department of Physics}, \orgname{Indian Institute of Technology Gandhinagar},  \city{Palaj}, \postcode{382 355}, \state{Gujarat}, \country{India}}

\affil[5]{\orgdiv{School of Fundamental Physics and Mathematical Sciences}, \orgname{Hangzhou Institute for Advanced Study, UCAS},  \city{Hangzhou}, \postcode{310 024},  \country{China}}

\affil[6]{\orgname{University of Chinese Academy of Sciences},  \city{Beijing}, \postcode{100 190}, \country{China}}

\abstract{In this paper, we study the evolutions of a self-gravitating cloud of bosonic dark matter with finite angular momentum and self-interaction. This is achieved by using the sixth-order pseudospectral operator splitting method to solve the system of nonlinear Schrödinger and Poisson equations. The initial cloud is assumed to have mass density randomly distributed throughout three-dimensional space. The dark matter particles in the initial cloud are in the kinetic regime, i.e., their de Broglie wavelength is much smaller than the halo size.

It is shown that Bose stars are indeed formed in the numerical simulation presented here. The presence of angular momentum and self-interaction in the initial cloud can significantly influence the star formation time in a non-trivial fashion. 

Furthermore, the plots of vorticity magnitude profile after the star formation time indicates that the formed star may not have any intrinsic angular momentum for the cases when the self-interaction among the particles is either negligible or attractive. These results are in agreement with the earlier analytical studies of an isolated rotating Bose star. However, for the case of repulsive self-interaction, the vorticity magnitude analysis show a possibility that the star formed in the numerical simulations may possess intrinsic angular momentum. It is also shown that the average mass and radius diagrams of the star are strongly influenced by the presence of angular momentum in the initial cloud.
}




\maketitle

\section{Introduction}\label{sec1}

Although the idea of Bose star dates back to several decades, the recent advances in the observational astronomy can provide an unique opportunity for detecting them. For example, the detection of the gravitational-wave signal GW190521 by the Advanced LIGO/Virgo collaboration \cite{Abbott} could be due to the merger of two boson stars \cite{Bustillo}.The idea of a Bose star is closely related to the concept of geons, first introduced by Wheeler in 1955 \cite{Wheeler}. Geons are considered to be macroscopic confined states of electromagnetic or gravitational radiation, held together by the gravitational field created by their own energy density. The notion of a boson star, formed by the scalar degree of freedom present in fundamental theories of physics undergoing gravitational collapse, was first introduced by Kaup in 1968 \cite{Kaup} and Ruffini and Bonazzola in 1969 \cite{RUFFINI}. Such a star can be regarded as a macroscopic state created by the balance between the quantum pressure of the scalar field due to the Heisenberg uncertainty principle and the force of self-gravity. There has been significant progress in finding solutions of the Einstein-Klein-Gordon equations for various situations in the literature \cite{Visinelli:2021uve}.  According to the standard model of cosmology, known as the $\Lambda$CDM model, cold dark matter constitutes approximately 83\% of the total matter in the Universe. Various candidates for dark matter, such as the QCD-axion \cite{PRESKILL1983127, Kolb, Turner, Hogan:1988mp}  and fuzzy cold dark matter \cite{Hu}, are described by the scalar degree of freedom. Hence, it is plausible that the Universe contains numerous Bose star-like objects. Moreover, gravitational bound state
structures formed by bosonic dark matter may help in explaining the missing satellite problem faced by $\Lambda$CDM model[for example see \cite{Levkov}].
There has been several studies that explore the connection between Bose stars and phenomena like fast radio bursts or the detection of axion stars \cite{Panin, Pshirkov:2016bjr, Raby:2016deh, Buckley}.

It should be noted that in the literature, not much attention has been paid to the issue of boson star formation under realistic scenarios. However, some interesting works have been published in recent years that focus on this issue \cite{Levkov, Chen2021, Veltmaat}.
In \cite{Levkov}, the authors study the gravitational collapse of a randomly distributed, non-interacting cloud of scalar bosons in the kinetic (low-density) regime. The kinetic regime is characterized by the de Broglie wavelength $(mv)^{-1}$ and time scale $mv^2$ of the particles, which satisfy the following conditions:

\begin{equation}
	mvR \gg 1 \quad \text{and} \quad mv^2 \tau_{\text{gr}} \gg 1  \label{eq1}
\end{equation}

Here, $m$ is the mass of the dark matter particle and $v$ is its typical velocity. $R$ represents the halo size, and $\tau_{\text{gr}}$ is the condensation time at which the star formation process begins. The authors estimate the value of $\tau_{\text{gr}}$ using a heuristic argument. After the time $\tau_{\text{gr}}$, numerical simulations show that the time evolution of the scalar field exhibits strong fluctuations during star formation, and the mass of the star continues to grow on average. The authors argue that their study can also help in solving the missing satellite problem \cite{Levkov}.

The role of self-interaction among the bosonic dark matter particles in the kinetic regime was analyzed in \cite{Chen2021}. This study shows that self-interaction in the dark matter cloud can influence the mass and compactness of the Bose star. Simulation of ultralight dark matter halo formation has been carried out in \cite{Veltmaat}. The study demonstrates a solitonic core and confirms the core-halo mass relations.

However, it should be emphasized that gravitational shearing can naturally induce rotation in star-forming clouds. For instance, velocity gradients were observed across the molecular clouds of the M33 galaxy, revealing their rotation \cite{larson, seigar}. Thus, considering the ubiquity of rotation in star-forming clouds, we investigate the formation and evolution of a Bose star when the initial dark matter distribution possesses finite angular momentum. Additionally, we explore the role of self-interaction in the star formation process while the initial dark matter density is randomly distributed throughout space, satisfying the conditions associated with the kinetic regime. In this work, we consider the self-interacting scalar dark matter field described by a Gross-Pitaevskii-type equation. First, we would like to note that there are existing studies in the literature that discuss related issues. In the reference \cite{Gual}, numerical solutions of the Einstein-Klein-Gordon and Einstein-Proca equations are considered to investigate the formation process of a Bose star. The initial field density profiles are examined with and without axisymmetry. It is demonstrated that in such cases, the scalar star is always unstable and loses all its angular momentum, resulting in the formation of a Saturn-like ring around the scalar star. However, the authors do not take into account any self-interaction. Moreover, in the Proca case, it was found that it might be possible to have a rotating stable star.

In the references \cite{Dmitriev, Siemonsen}, the authors consider the initial configuration as an isolated boson star with finite angular momentum and study their stability. The authors show that rotating Bose stars are always unstable when the self-interaction between their constituent particles is either negligible or attractive, leading to the loss of all angular momentum. Interestingly, when the self-interaction is repulsive, it is possible to have a rotating star.
In reference \cite{Dmitriev}, an analytical expression for the growth rate of the instability was obtained. The dark matter field in this reference is described by a Gross-Pitaevskii-type equation. On the other hand, the reference \cite{Siemonsen} considers the Klein-Gordon field.

Therefore, we believe that the approach chosen in our present paper will provide useful insights into the formation process of Bose stars. Additionally, it will allow us to compare our results with the related works discussed here.

This work is divided into the following sections: In Section \ref{sec2}, we discuss the time evolution of boson particles for various cases, including attractive/repulsive self-interaction and angular momentum. In Section \ref{sec3}, we present the results for the aforementioned cases. Finally, we summarize and conclude the results in Section \ref{sec4}. Throughout the paper, we use natural units where $c$, $k_B$, and $\hbar$ are set to 1. Here, $c$ represents the speed of light, $k_B$ is the Boltzmann constant, and $\hbar$ is the reduced Planck constant. Boldface quantities indicate three-dimensional vectors.

\section{Evolution of the Bose star}\label{sec2}

In the non-relativistic limit and for a high mean occupation number, the system can be represented by a classical scalar field, $\psi (\bm r,t)$, whose time evolution can be described by the Gross-Pitaevskii-Poisson equations \cite{Chen2021, Chavanis, Eby:2015hsq} :

\begin{alignat}{2}
i\frac{\partial \psi}{\partial t} & = -\frac{1}{2m}\nabla^2\psi + m\Phi\psi + g|\psi|^2\psi, \label{eq2} \\
\nabla^2\Phi & = 4\pi Gm(|\psi|^2 - n), \label{eq3}
\end{alignat}

Here, $\Phi\equiv\Phi(\bm r,t)$ is the gravitational potential, $G$ is the universal gravitational constant, and $n$ is the mean particle density of the boson gas. The parameter $g$ represents the self-interaction coupling constant, where positive values correspond to repulsive self-interaction and negative values correspond to attractive self-interaction. The case $g=0$ corresponds to a non-interacting scenario. For a box with a size $L$ and a total number of particles $N$, the number density is given by $n=N/L^3$. The above equations \eqref{eq2} and \eqref{eq3} can be written in dimensionless form using the following scaling: $r=(1/mv_0)\tilde{r}$, $t=(1/mv_0^2)\tilde{t}$, $\Phi=v_0^2\tilde{\Phi}$, $\psi=(v_0^2\sqrt{m/4\pi G})\tilde{\psi}$, and $g=(4\pi G/v_0^2)\tilde{g}$:

\begin{alignat}{2}
i\frac{\partial}{\partial\tilde{t}}\tilde{\psi} & = -\frac{1}{2}\tilde{\nabla}^2\tilde{\psi} + \tilde{\Phi}\tilde{\psi} + \tilde{g}|\tilde{\psi}|^2\tilde{\psi}, \label{eq4} \\
\tilde{\nabla}^2\tilde{\Phi} & = |\tilde{\psi}|^2 - \tilde{n}, \label{eq5}
\end{alignat}

Here, $\tilde{n} = \dfrac{4\pi G}{mv_0^4} n$, and $N=\int d^3r|\psi|^2$ can be expressed as $N = \dfrac{v_0}{4\pi Gm^2}\tilde{N}$, where $v_0$ is the reference velocity \cite{Chen2021}.

Next, numerical solutions of the above equations are obtained using the pseudospectral operator splitting method. This is done by utilizing the code \texttt{AxioNyx}\footnote{\href{https://github.com/axionyx}{https://github.com/axionyx}} \cite{Schwabe}. However, we needed to modify the code to incorporate the effect of self-interaction in the method.

To proceed further, we need to specify the initial condition for $\tilde{\psi}$ at time $\tilde{t}=0$. First, consider the case when there is no angular momentum carried by the function $\tilde{\psi}$. In this case, the wave function is assumed to be 'Gaussian' and can be written in Fourier space as:

\begin{equation}
\tilde{\psi}_p = \frac{\tilde{N}^{1/2}}{\pi^{3/4}}e^{-\tilde{p}^2/2} e^{i\alpha_p},\label{eq6}
\end{equation}

Here, $\alpha_p$ represents a random phase distributed over the Fourier space between the numbers 0 and $2\pi$. To generate a randomly distributed matter density distribution in space, we transform to the coordinate space to obtain the profile of amplitude $\tilde{\psi}(\tilde{x},\tilde{y},\tilde{z}, \tilde{t}=0)$. This prescription is the same as that considered in reference \cite{Levkov}.

The angular momentum in the initial wavefunction is incorporated using the following prescription: First, consider the function

\begin{alignat}{2}
\tilde{f}(\tilde{\bm r},\tilde{t}=0) =  e^{-\tilde{\bm r}^2/2} e^{il\tan^{-1}(\tilde{y}/\tilde{x})},\label{eq7}
\end{alignat}

\begin{figure*}
	\centering
	\begin{tabular}{@{}c@{}}
		\includegraphics[width=14cm,height=5cm]{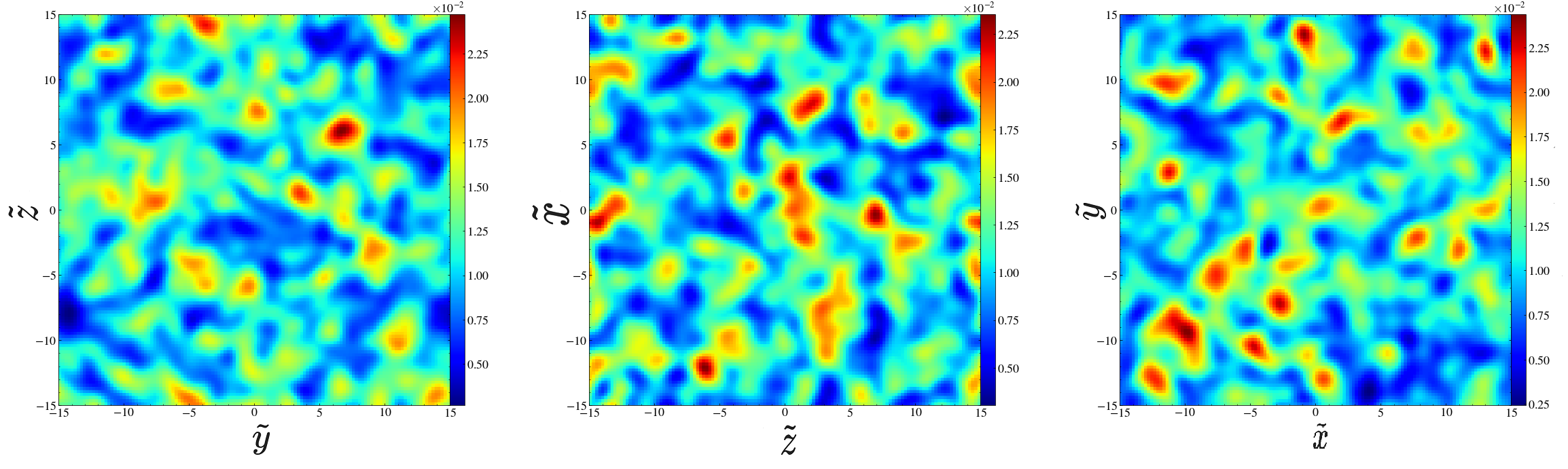} \\[\abovecaptionskip]
		%\small (a) 
	\end{tabular}	
	
%	\begin{tabular}{@{}c@{}}
%		\includegraphics[width=18cm,height=7cm]{l5} \\[\abovecaptionskip]
%		%\small (b)
%	\end{tabular}
	
	\begin{tabular}{@{}c@{}}
		\includegraphics[width=14cm,height=5cm]{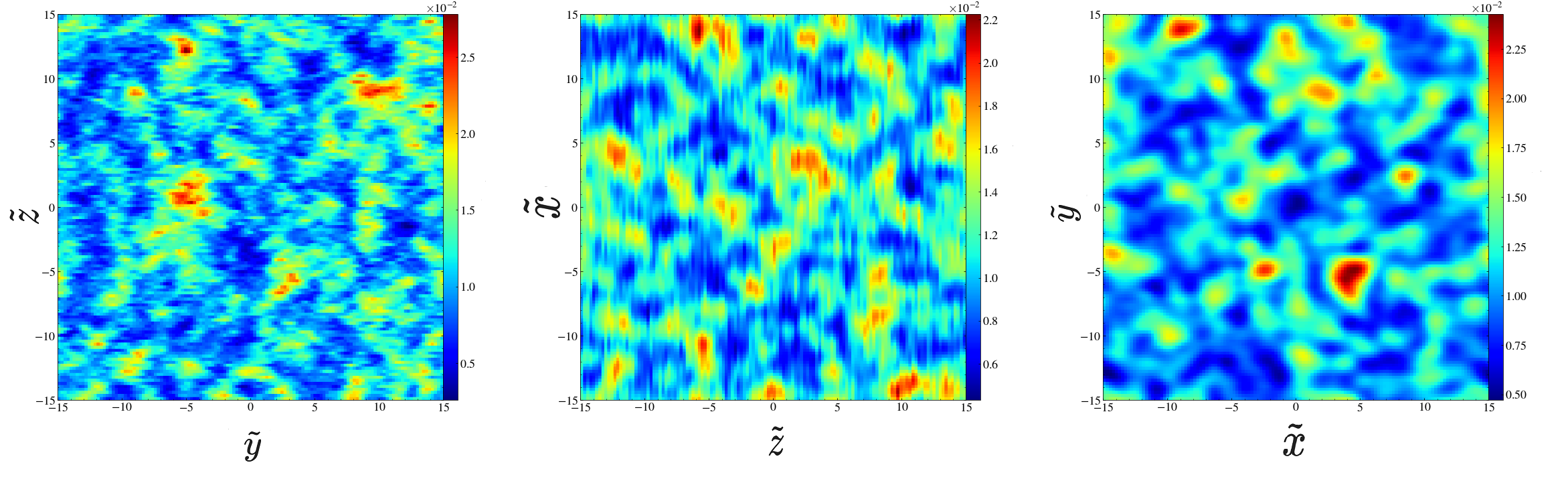} \\[\abovecaptionskip]
		%\small (b)
	\end{tabular}

	\caption{The initial snapshots of density profiles of $|\tilde{\psi}|^2$ at the initial time ($\tilde t=0$) in $\tilde y \tilde z$, $\tilde z \tilde x$ and $\tilde x\tilde y$ planes. The top panel from left to right describes the initial cloud density for $\mathcal{\tilde L_{\rm tot}}=0$. The bottom panel respectively correspond to $\mathcal{\tilde L_{\rm tot}}=5$ case.}\label{plot1}
\end{figure*}

This function is then transformed to Fourier space to obtain $\tilde{f}(\bm p)$. The initial wavefunction in momentum space can be obtained as $\tilde{\psi}(\bm p) = \tilde{f}(\bm p) \times e^{i\alpha_p}$, where $\alpha_p$ is the random phase as defined above. The input wavefunction $\tilde{\psi}(\tilde{x},\tilde{y},\tilde{z}, \tilde{t}=0)$ is generated by transforming $\tilde{\psi}(\bm p)$ back to coordinate space. For all the numerical results presented in this paper, we use the aforementioned prescription. It is possible to modify the prescription described by equation \eqref{eq7}; in fact, we have tried a couple of different prescriptions to introduce angular momentum in $\tilde{\psi}(\tilde{x},\tilde{y},\tilde{z}, \tilde{t}=0)$, but it does not change the value of $\tilde{\tau}_{\text{gr}}$ for a given value of angular momentum. The qualitative behavior of the numerical results remains the same. Fig.\eqref{plot1} shows the initial density profile $|\tilde{\psi}(\tilde{x},\tilde{y},\tilde{z}, \tilde{t}=0)|^2$ for different values of angular momentum in different planes.

\begin{figure*}
	\centering
		\includegraphics[width=14cm,height=8cm]{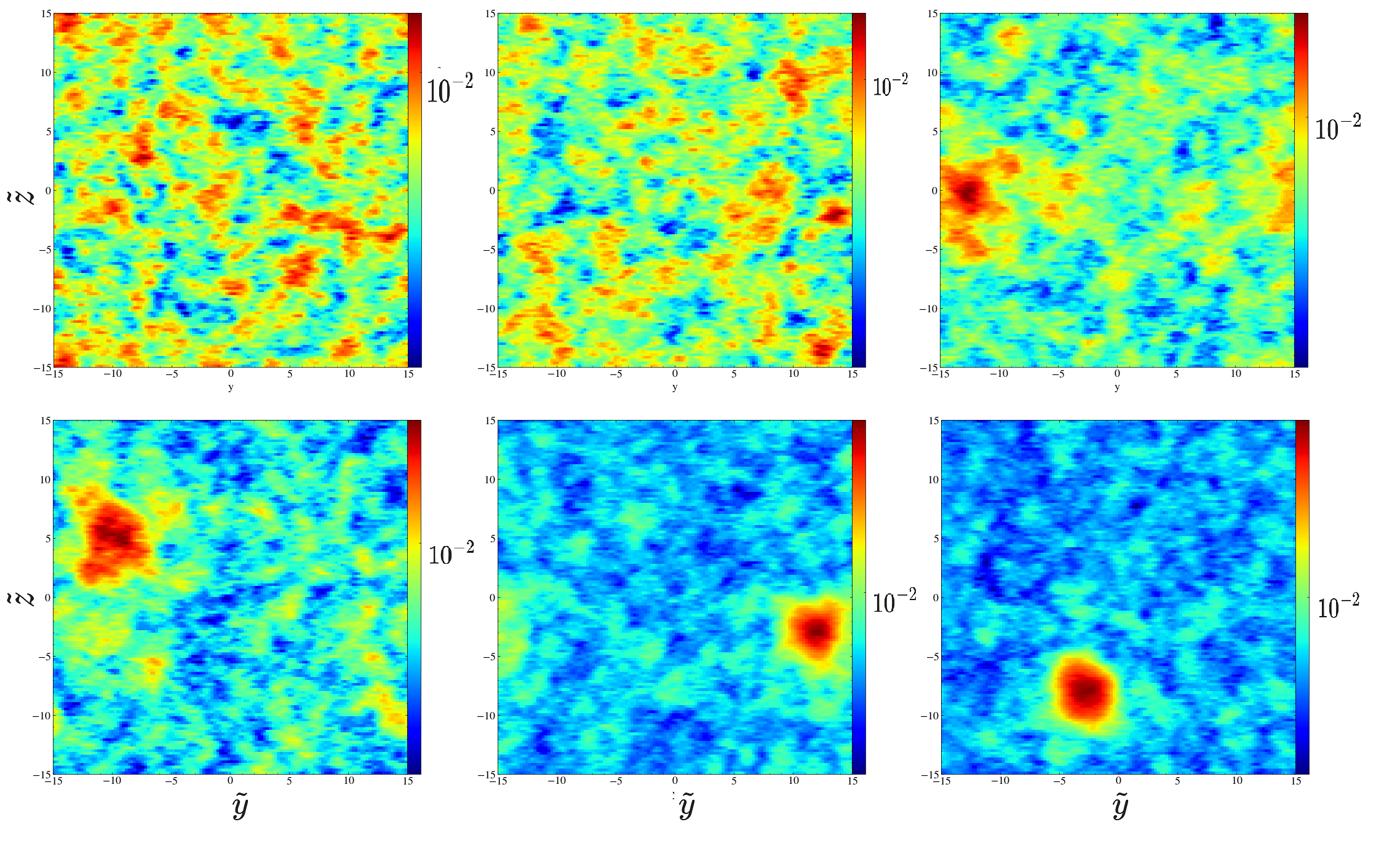}
		%\small (a) 	
	\caption{Snapshots of $|\tilde{\psi}|^2$ at different values of time are shown in $\tilde{z}\tilde{y}$-plane. All the plots represent the case when the initial cloud has total-angular momentum $\mathcal{\tilde L_{\rm tot}}=5$ 
		and no self-interaction $\tilde{g}=0$.  The gravitational condensation time $\tilde{\tau}_{gr}$ for this case is around 15600. The plots in the upper panel from the left to right respectively correspond to time $\tilde{t}=0,\, 0.5\tau_{gr},\, \& 1\tau_{gr}$. Three plots in the bottom correspond to $1.15\tilde{\tau}_{gr},\,1.5\tilde{\tau}_{gr}\,\& 1.8\tilde{\tau}_{gr}$.} \label{plot2}
\end{figure*}

To calculate the total angular momentum carried by the initial scalar field, we consider the stress-energy tensor \cite{Weinberg:1972kfs}:

\begin{alignat}{2}
\widetilde{T}^{0i} = \frac{i}{2} \left(\tilde{\psi}\tilde{\partial}_i\tilde{\psi}^* - \tilde{\psi}^*\tilde{\partial}_i\tilde{\psi}\right).\label{eq8}
\end{alignat}

The total angular momentum in the $\tilde{x}$, $\tilde{y}$, and $\tilde{z}$ directions, in terms of the stress-energy tensor, can be defined as \cite{Weinberg:1972kfs}:

\begin{alignat}{2}
\tilde{J}^{ij} = \int(\tilde{x}^i\widetilde{T}^{j0} - \tilde{x}^j\widetilde{T}^{i0})d^3\tilde{x}.\label{eq9}
\end{alignat}

For the considered initial distribution in equation \eqref{eq7}, the total angular momentum in the $\tilde{x}$ and $\tilde{y}$ directions ($\tilde{J}^{2,3}$ and $\tilde{J}^{3,1}$, respectively) is zero. Therefore, the total angular momentum of the system is $\mathcal{\tilde{L}_{\text{tot}}} = \tilde{J}^{1,2} = l\tilde{N}$. It is important to note that once a star is nucleated, it will be performing random motion. Therefore, even in the case with $l=0$, the star will have some orbital angular momentum. In cases when the initial cloud has finite angular momentum, it is hard to study angular momentum transfer from the cloud to the star because of the presence of orbital and also the possibility of having intrinsic angular momentum. In this case, we find it useful to study how the vorticity magnitude is distributed outside the star formation region. However, this approach does not allow us to quantify angular momentum transfer from the surrounding cloud to the star. To determine the vorticity of the system, we introduce the 'velocity' ${\bf v} = \bm{\mathcal{J}}/\rho$, where $\bm{\mathcal{J}}$ represents the current density and $\rho=|\psi|^2$. The current density $\bm{\mathcal{J}}$ is defined from equation \eqref{eq1} as $\bm{\mathcal{J}} = \frac{i}{2}(\psi{\bm{\nabla}}\psi^* - \psi^*{\bm{\nabla}}\psi)$. From this, one can define the vorticity as ${\bf\Omega} = {\bm{\nabla}}\times{\bf v}$, and the vorticity magnitude as $\Omega = \sqrt{\Omega_x^2 + \Omega_y^2 +\Omega_z^2  }$.

In the subsequent discussion, the magnitude of vorticity in different planes has been used to understand how the angular momentum is distributed throughout the space during the process of star formation.

It is important to emphasize that the presence of attractive self-interaction can also allow for the possibility of forming bound states without the self-gravity playing a significant role \cite{Chen2021}. Thus, for a gravitational bound state to be produced in the numerical solution, it is expected that the self-coupling term remains small or comparable to the self-gravity term. Furthermore, the self-interaction term can influence the typical size of the structure produced by self-gravity, even when there is no angular momentum in the rotating cloud. In the kinetic regime, one may not be able to use the hydrodynamic formalism to estimate the Jeans length. Therefore, in the present case, the following arguments can be used to estimate the typical size of the gravitational structure and provide a bound on the coupling constant.

If we replace all spatial derivative terms in the GPP equations (i.e., equations \eqref{eq3} and \eqref{eq4}) with $1/\tilde{L}$, where $\tilde{L}$ is the length scale, we get:

\begin{equation}
i\frac{\partial \tilde{\psi}}{\partial t} = \left[\frac{1}{2\tilde{L}^4} - \left(|\tilde{\psi}|^2 - \tilde{n}\right) + \tilde{g}\frac{|\tilde{\psi}|^2}{\tilde{L}^2}\right] \tilde{\psi}\tilde{L}^2.\label{eq10}
\end{equation}

The system would collapse under gravity. If there are no particles in the surroundings, then a stationary state can be reached, i.e., $i\frac{\partial \tilde{\psi}}{\partial t} \sim \omega\tilde{\psi}$. By relabeling $\tilde{L}$ with the symbol $\tilde{L}_B$, where the subscript $B$ indicates a bound state, we have:

\begin{equation}
\omega\tilde{\psi} = \left[\frac{1}{2\tilde{L}_B^4} - \left(|\tilde{\psi}|^2 - \tilde{n}\right) + \tilde{g}\frac{|\tilde{\psi}|^2}{\tilde{L}_B^2}\right] \tilde{\psi}\tilde{L}_B^2.\label{eq11}
\end{equation}

Here, the term with $\tilde{n}$ is neglected in comparison with the $|\tilde{\psi}|^2$ term. The first and second terms in the square bracket correspond to the quantum pressure and self-gravity effects, respectively, while the third term is the contribution from self-interaction. For a gravitationally bound state to be formed in the collapse, we require that the second term in the square bracket remain larger than the term with self-coupling, leading to the following condition:

\begin{equation}
|\tilde{g}\tilde{L}_B^{-2}| < 1. \label{eq12}
\end{equation}

Thus, we can estimate the size of the bound structure by solving the biquadratic equation:

\begin{equation}
\tilde{L}_B^2 = -\frac{\omega - \tilde{g}|\tilde{\psi}|^2}{2|\tilde{\psi}|^2} + \sqrt{\frac{(\omega - \tilde{g}|\tilde{\psi}|^2)^2 + 2|\tilde{\psi}|^2}{4|\tilde{\psi}|^4}}, \label{eq13}
\end{equation}

where we have ignored the negative square root solution as it could result in $\tilde{L}_B^2 < 0$. Since $\omega < 0$ for gravitationally bound states, from the above equation, it can be seen that for attractive self-interaction (i.e., $\tilde{g} < 0$), the formed structure is smaller in size compared to the cases with $\tilde{g} > 0$ and $\tilde{g} = 0$.

The restriction on the values of the coupling constant can also be obtained using a different set of arguments \cite{Jiajun}. Here, the timescale $\tau_{\text{self}}$ for obtaining a bound state without self-gravity is given by, $\tau_{\text{self}} = \dfrac{4 \sqrt{2} mv^2}{3 n^2 g^2 \pi}$, when there is an attractive interaction between the constituent particles of the initial cloud. For the case when both self-interaction and self-gravity are present, their combined effect can give rise to a new timescale $\tilde{\tau}_{\text{eff}}$ for structure formation, which is given by:

\begin{equation}
\tilde{\tau}_{\text{eff}} = \frac{{\tilde{\tau}}_{\text{gr}}\tilde{\tau}_{\text{self}}}{\tilde{\tau}_{\text{gr}} + \tilde{\tau}_{\text{self}}}.\label{eq14}
\end{equation}

Thus, when the condition $\tilde{\tau}_{\text{gr}} < \tilde\tau_{\text{self}}$ is satisfied, self-gravity can play a dominant role in structure formation.

%%%%%%%%%%%%%%%%%%%%%%%%%%%%%

%%%%%%%%%%%%%%%%%%%%%%%%%%%%%

\section{Results and Discussion}\label{sec3}
The density profiles of the initial wave-functions in the $yz$, $zx$, and $xy$ planes for various values of angular momentum $\mathcal{L}_{\text{tot}}=0$ are shown in Figure \eqref{plot1}. In the absence of angular momentum or self-coupling in the initial boson cloud, the only important timescale is the gravitational condensation time described by the formula given in \cite{Levkov}:

\begin{equation}
\tau_{\text{gr}} = \frac{b \sqrt{2}}{12 \pi^3} \frac{mv_0^6}{ G^2 n^2 \Lambda}.\label{eq15}
\end{equation}

where $n$ is the boson number density and $\Lambda = \log(mv_0L)$, with $\tilde{L} = mv_0 L$ being the box length used in the numerical simulation. It should be noted that the complete determination of the parameter $b$ is not possible, but it can be of the order of unity \cite{Levkov}. In the subsequent section, we will discuss the influence of angular momentum and self-interaction on the parameter $b$. Using dimensionless units, we can write the gravitational condensation time as:
\begin{equation}
\tilde{\tau}_{\text{gr}} = \frac{b \sqrt{2}}{12 \pi^3} \frac{1}{\tilde{n}^2 \log(\tilde{L})}..\label{eq16}
\end{equation}

To ensure the validity of the numerical results obtained using the modified code, we have employed the following checks:

\textbf{1)} In the absence of angular momentum and self-interaction, the code produces the results from \cite{Levkov} and \cite{Schwabe}, i.e., the estimated value of $\tilde{\tau}_{\text{gr}}$ is in agreement with the numerical values obtained by the code. Also, the power spectrum of the wave-function $F(\omega, t)$ (defined below) develops a peak at the negative frequency value after $\tilde{t} > \tilde{\tau}_{\text{gr}}$.

\textbf{2)} The total energy of the system is conserved for the system of equations \eqref{eq4} \& \eqref{eq5}. To check energy conservation, the total energy is calculated for the initial data and compared with the total energy at each subsequent step. If at any time the total energy of the system exceeds the initial energy by more than 5\%, we do not run the code any further.

Next, keeping the available computational resources in mind, we choose $\tilde{N} = 10$, $\tilde{L} = 30$, and a resolution of $\tilde{L}^3/128^3$. The typical value of $\tilde{\tau}_{\text{gr}}$ is around $8 \times 10^3$, and further increasing the resolution can significantly increase computational time. Also, including repulsive interactions or higher angular momentum can further increase the condensation time $\tilde{\tau}_{\text{gr}}$. It should be noted that for this resolution, the continuum limit can be achieved, as shown in \cite{Levkov}(also see \cite{Chen2021}). Here, the authors use the exceptionally stable sixth order pseudospectral operator splitting method which is suitable for a long statistical simulations. In this work we employ the same technique to study time evolution of a self-gravitating cloud of a dark-matter in the kinetic regime.

For the chosen values of $\tilde{N}$ and $\tilde{L}$, the highest allowed value of $|\tilde{g}|$ for the formation of a gravitationally bound state is 4.56, as given by the bound condition \eqref{eq12}. We have checked if the numerical results show the process of star formation for  $|\tilde{g}|>4.56$ ; however, we do not find star formation occurring for either attractive or repulsive self-interaction within the time allowed by our energy conservation test. Moreover, if we reduce the highest values of $|\tilde{g}|$ by an order of magnitude, the numerical solutions are very similar to the case with $\tilde{g} = 0$. Thus, in this work, we present the results for the following three values of the coupling constant: $\tilde{g} = -4.56$, 0.0, and +4.56. The data generated by the code is analyzed using \texttt{The yt Project} \cite{Turk_2011}.

Here, we would like to note that recently in \cite{Levkov} and \cite{Siemonsen}, the authors studied the stability of an isolated rotating Bose star. It was demonstrated that when the self-interaction among the constituent particles is attractive or negligibly small, the star becomes unstable for any value of angular momentum. The instability results in the shedding of particles (including angular momentum) in the region outside the star, leading to the formation of a Saturn ring-like structure around the star. The time scale associated with the instability of a rotating boson star \cite{Dmitriev} was estimated using the growth rate given by:

\begin{equation}
\tau_{\text{ins}} = \frac{100}{2.2} \frac{l^2}{\alpha_l} \frac{1}{m^3 G^2 M_s^2}..\label{eq17}
\end{equation}

where $l$, $M_s$, and $\alpha_l$ represent the values of angular momentum, the mass of the star, and an angular-momentum dependent parameter, respectively. However, the authors demonstrate that when the self-interaction between the constituent particles is repulsive, it is possible to have a stable Bose star configuration.

Thus, it would be rather interesting to see whether the Bose stars formed in our numerical results are rotating or not. First, we would like to emphasize that in the stability analysis by \cite{Dmitriev}, the initial configuration is assumed to be an isolated boson star. In our work, star formation occurs after a time $\tilde{\tau}_{\text{gr}}$ from the initial state with a randomly distributed density over the entire space. To understand how the instability influences the presented numerical results, let us consider the ratio ${\tilde{\tau}_{\text{ins}}}/{\tilde{\tau}_{\text{gr}}}$. Using the values of $\tilde{n}$ and $\tilde{L}$, the ratio ${\tilde\tau_{\text{ins}}}/{\tilde\tau_{\text{gr}}} \sim 10^{-2}$. Furthermore, it should be noted that we have used the value of $\tilde{\tau}_{\text{gr}} \sim 8000$. The typical range of $\tilde{\tau}_{\text{gr}}$ is around 8000-25000 for $\tilde{g} \geq 0$. Thus, one can only decrease ${\tilde{\tau}_{\text{ins}}}/{\tilde{\tau}_{\text{gr}}}$ by increasing $\tilde{\tau}_{\text{gr}}$. Since ${\tilde\tau_{\text{ins}}}/{\tilde\tau_{\text{gr}}} \leq 10^{-2}$, we believe that any possible instability would have time to saturate.

To check when the star condensation begins, we employ the following three tests:
\begin{itemize}
    \item The first test involves studying the time evolution of the maximum amplitude $|\tilde{\psi}(\tilde{x}, \tilde{t})|$. The typical value for the initial maximum amplitude is $|\tilde{\psi}(\tilde{x}, \tilde{t})| \sim 2 \times 10^{-2}$. The sudden rise in the maximum amplitude could be due to gravity or the self-interaction producing a bound state. This provides us with an estimate of $\tilde{\tau}_{\text{eff}}$ for $\tilde{g} \leq 0$ or $\tilde{\tau}_{\text{gr}}$ when $\tilde{g} > 0$ for repulsive interaction.
    \item The second test involves studying the power spectrum $F(t,\omega)$ of $\psi(x,t)$ \cite{Levkov}, defined as the Fourier transform of the correlator:

\begin{equation}
F(\omega, t) = \int \frac{dt_1}{2 \pi} d^3\bm{x}\ \psi^*(t,\bm {x}), \psi(t+t_1,\bm{x}), e^{i \omega t_1 - t_1^2/\tau_1^2}\label{eq18}
\end{equation}

Around the time when gravitational condensation occurs, $F(\omega, t)$ starts developing a peak at $\omega_s < 0$, indicating the formation of a gravitationally bound state.
    \item The third test involves checking if the formed structure is gravitationally bound by satisfying the condition $|\tilde{g}\tilde{L}_B^{-2}| < 1$.

\end{itemize}
These tests allow us to determine the onset of star condensation and verify the formation of a gravitationally bound structure.

\begin{figure*}
	\centering
	\begin{subfigure}{6cm}
		\includegraphics[width=6cm,height=8.5cm]{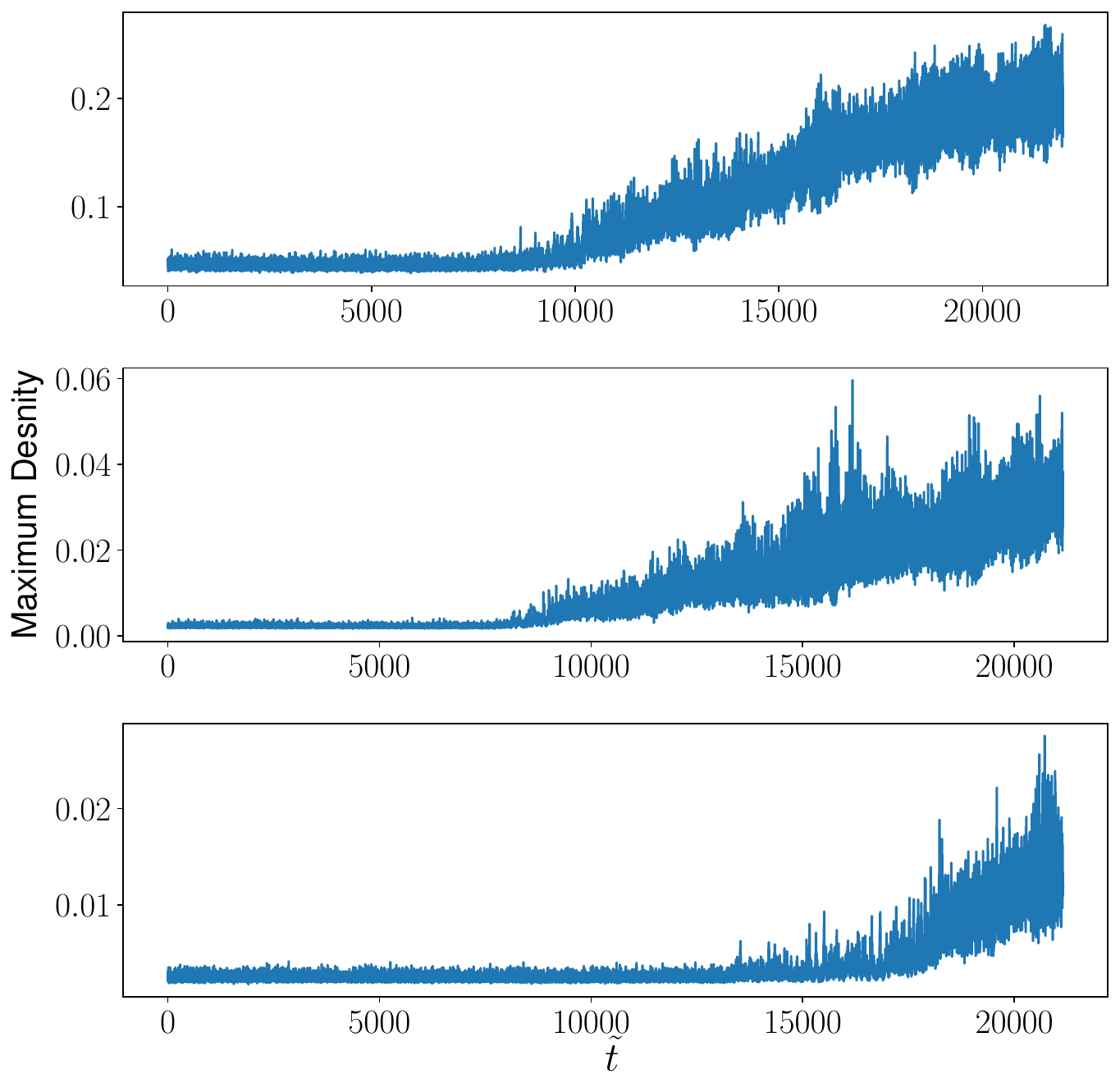}
		\caption{}
		\label{plot3a}
	\end{subfigure}
	\hspace{1cm}
	\begin{subfigure}{6cm}
		\includegraphics[width=6cm,height=8.5cm]{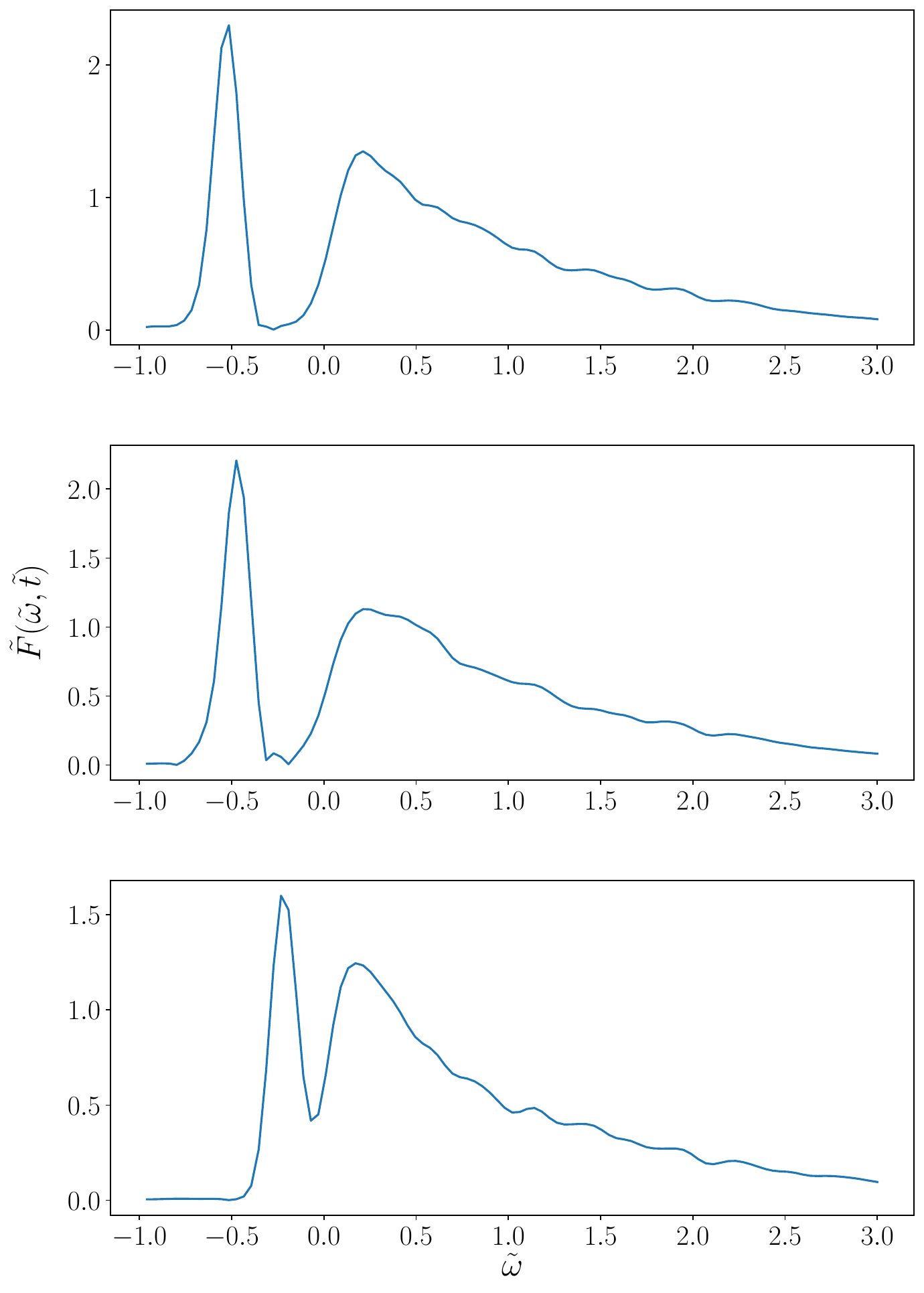}
		\caption{}
		\label{plot3b}
	\end{subfigure}
	\caption{Figure \ref{plot3a} represents the time evolution of the maximum density $|\psi|^2$ for different values of initial angular momentum of the bosonic field for $\tilde{g}=0$. Figure \ref{plot3b} represents the energy spectrum, $\tilde{F}(\tilde{\omega})$, as a function of $\tilde{\omega}$. In both figures, $\mathcal{\tilde{L}_{\rm tot}}$ varies from top to bottom as 0, 3, and 5. In figure \ref{plot3b}, all plots are obtained at $\tilde{t} \sim 21000$.}
	\label{plot3}
\end{figure*}

Figure \eqref{plot3a} shows the evolution of the maximum density, $\max|\tilde{\psi}|^2$, with time $\tilde{t}$. The plots from top to bottom correspond to cases with $\mathcal{\tilde{L}_{\text{tot}}} = 0$, $3$, and $5$. Initially, $\max|\tilde{\psi}|$ remains constant and small for a long time $\tilde{t} \sim \tilde{\tau}_{\text{gr}}$, and then there is a sudden rise in its value. It should be noted that the mean value of density continues to rise monotonically for $\tilde{t} > \tilde{\tau}_{\text{gr}}$ until the convergence test stops the code around $\tilde{t} \sim 21000$. Ideally, the growth in density should stop at some point, but due to computational limitations, it is difficult to observe this in the numerical computation. Previous studies related to Bose star formation, such as \cite{Levkov} and \cite{Chen2021}, have also not observed saturation in density in the kinetic regime. Additionally, the inclusion of angular momentum in the initial amplitude does not help in achieving saturation.

Figure \eqref{plot3b} describes the behavior of the spectrum $\tilde{F}(\tilde{\omega}, \tilde{t})$ as a function of frequency at $\tilde{t} \sim 21000$, which is larger than the condensation time $\tilde{\tau}{\text{gr}}$ for all three cases. The figure clearly shows the presence of a prominent peak in the region where the frequency has negative values for all different values of $\mathcal{\tilde{L}}_{\text{tot}}$. It should be emphasized that this peak in  $\tilde{F}(\tilde{\omega}, \tilde{t})$ occurs soon after $\tilde{\tau}_{\text{gr}}$ and persists as long as the convergence test remains valid. This peak indicates the presence of a gravitationally bound state or a Bose star.

\begin{figure*}
	\includegraphics[width=14cm,height=4cm,trim={0 0 1cm 0},clip]{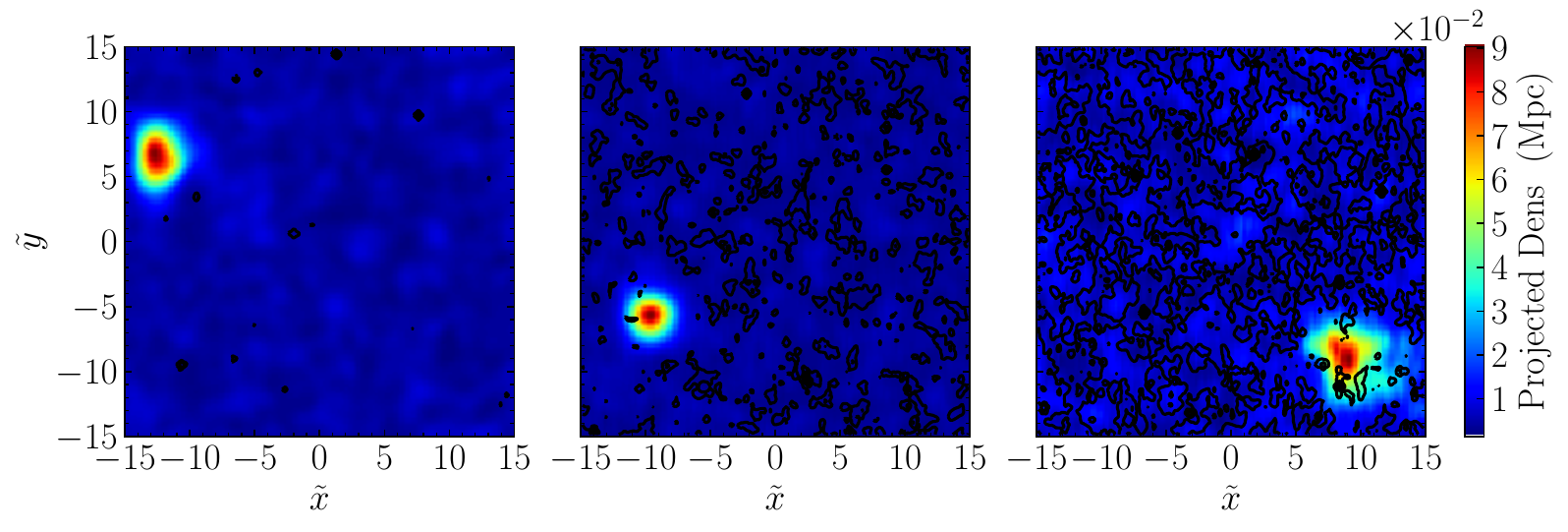}
	\caption{  Three snapshots of density profile $|\tilde{\psi}|^2$ together with vorticity magnitude in  $\tilde x\tilde y-$plane are shown at time  $\tilde{t} \sim $ 21000. This the case when there is no self-interaction.
		The first plot on the left corresponds to the case with no angular momentum ($\mathcal{\tilde L_{\rm tot}}=0$), while the second and the the third plots respectively describe the  cases when $\mathcal{\tilde L_{\rm tot}}=3$ and 5. The black contours represent vorticity magnitude.
	The high density region inside the redspots corresponds to the Bose star. The second and third plots shows the presence of vorticity outside the star formation region. In $\tilde x\tilde z-$plane \& $\tilde y\tilde z-$planes also vorticity found to be in the region outside the star.}\label{plot4}
\end{figure*}

To gain insight into the fate of the initial cloud's angular momentum, we have shown three snapshots of the density profile $|\tilde \psi|^2$ along with the vorticity magnitude in the $\tilde{x}\tilde{y}$-plane at a time around 21000 in Figure \eqref{plot4}. The vorticity magnitude is represented by the black contours. The leftmost plot corresponds to the case with $\mathcal{\tilde{L}}_{\text{tot}} = 0$, where there is no initial vorticity. In the other two plots with non-zero angular momentum values, most of the vorticity magnitude remains outside the star-forming region. This suggests that the stars formed in the cases with $\mathcal{\tilde{L}}_{\text{tot}} = 3$ and $\mathcal{\tilde{L}}_{\text{tot}} = 5$ may not possess intrinsic angular momentum. These findings align with the stability analysis presented in \cite{Dmitriev}.

The typical time interval for which the star persists after the condensation time is approximately (1.5-2.0)$\tilde{\tau}_{\text{gr}}$, with the specific value depending on $\mathcal{\tilde{L}}_{\text{tot}}$ and type of self-interaction. The instability of a rotating boson star in the absence of self-interaction is known, but the characteristic time for the instability to grow, $\dfrac{\tilde{\tau}_{\text{ins}}}{\tilde{\tau}_{\text{gr}}} \sim 10^{-2}$, is much shorter than (1.5-2.0)$\tilde{\tau}_{\text{gr}}$. Therefore, it is expected that the instability saturates within the observed time interval. Additionally, the presence of a peak in $\tilde F(\tilde \omega, \tilde t)$ at negative frequencies throughout the interval (1.5-2.0)$\tilde{\tau}_{\text{gr}}$ suggests that the star formed in the case with $\mathcal{\tilde{L}}_{\text{tot}} = 0$ is stable against rotational instability.

In the left panel of Figure \eqref{plot3a}, it can be seen that the maximum density starts to increase and fluctuate strongly after the condensation time. Therefore, the plots of total mass of the star $M(\tilde{t})$ versus time $\tilde{t}$ can exhibit growth, but the data also shows large fluctuations. Similarly, one would expect to observe fluctuations in the radius versus time plots as well. The theoretical calculations based on stationary boson stars predict that as the mass of the star increases, its radius shrinks \cite{Chavanis, Schiappacasse:2017ham}. This behavior has also been observed in numerical simulations \cite{Levkov}. Furthermore, it has been shown that if the mass and radius of a non-rotating Bose star increase simultaneously, they can be unstable \cite{Schiappacasse:2017ham}. In order to smooth out the fluctuations in the data, we employ a digital filter developed by Savitzky-Golay \cite{Savitzky}.Compared to other smoothing techniques(such as moving average, etc), the Savitzky-Golay filter stands out for its ability to simultaneously reduce noise and maintain the integrity of fine details present in the data. By fitting a local polynomial function to a sliding window of data points, the filter effectively captures the underlying trend while suppressing high-frequency noise components.

\begin{figure*}
	\includegraphics[width=14cm,height=8cm]{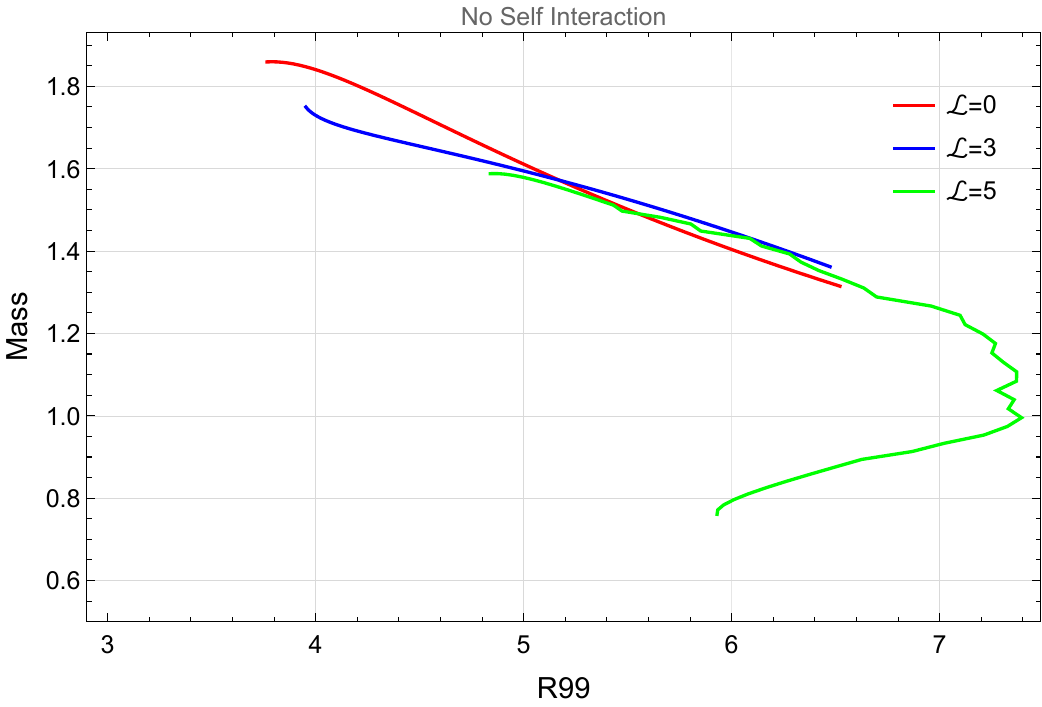}
	\caption{Mass–radius relation for Bose stars without any self-interaction for different values of $\mathcal{\tilde L_{\rm tot}}$}\label{plot5}
\end{figure*}

Figure \eqref{plot5} shows the $M-R$ relation, $R_{99}$ measures  distance from the centre at which 99\% mass of the star is located. It has been observed in our simulation that $2 R_{99}\,>\,L_B$. The red curve represents the case with no angular momentum in the initial cloud. It can be clearly seen that the star becomes more and more compact as its mass increases. Similar behavior is observed for the blue curve representing the case with $\mathcal{\tilde{L}}_{\text{tot}} = 3$. However, for the case with $\mathcal{\tilde{L}}_{\text{tot}} = 5$, there are two branches of solutions. In the lower part of the green curve, both the mass and radius of the star increase together. This branch may be unstable \cite{Schiappacasse:2017ham}, but the time scale of the instability may be much longer than $\tilde{\tau}_{\text{gr}}$, allowing the star to survive during the simulation time. The lowest mass on this diagram occurs around the time when the condensation just begins, and with the passage of time, the star moves towards the stable branch.

\subsection{Bose star formation in  presence of self-interaction} 
Next, we investigate Bose star formation in the presence of finite self-interaction. We examine the gravitational collapse of a self-interacting boson field for various values of the coupling constant $\tilde g$. As discussed earlier, star formation may not be possible for arbitrarily large values of the coupling strength. For $\tilde{N}=10$ and $\tilde{L}=30$, we determine that the maximum allowed value of $|\tilde{g}|$ is approximately 4.56. However, if we decrease the coupling constant by an order of magnitude, the results approach the case with no self-interaction. Therefore, in this section, we focus our analysis on a high value of the coupling constant.

\subsubsection{Attractive self-interaction}

In this case, it is possible to form bound structures due to the balance between the attractive force provided by the self-interaction and the quantum pressure. Hence, it is crucial to examine the role of self-gravity in the observed bound state. For this analysis, we consider $\tilde{g}=-4.56$.

\begin{figure*}
	\centering
	\begin{subfigure}{6cm}
		\includegraphics[width=6cm,height=8.5cm]{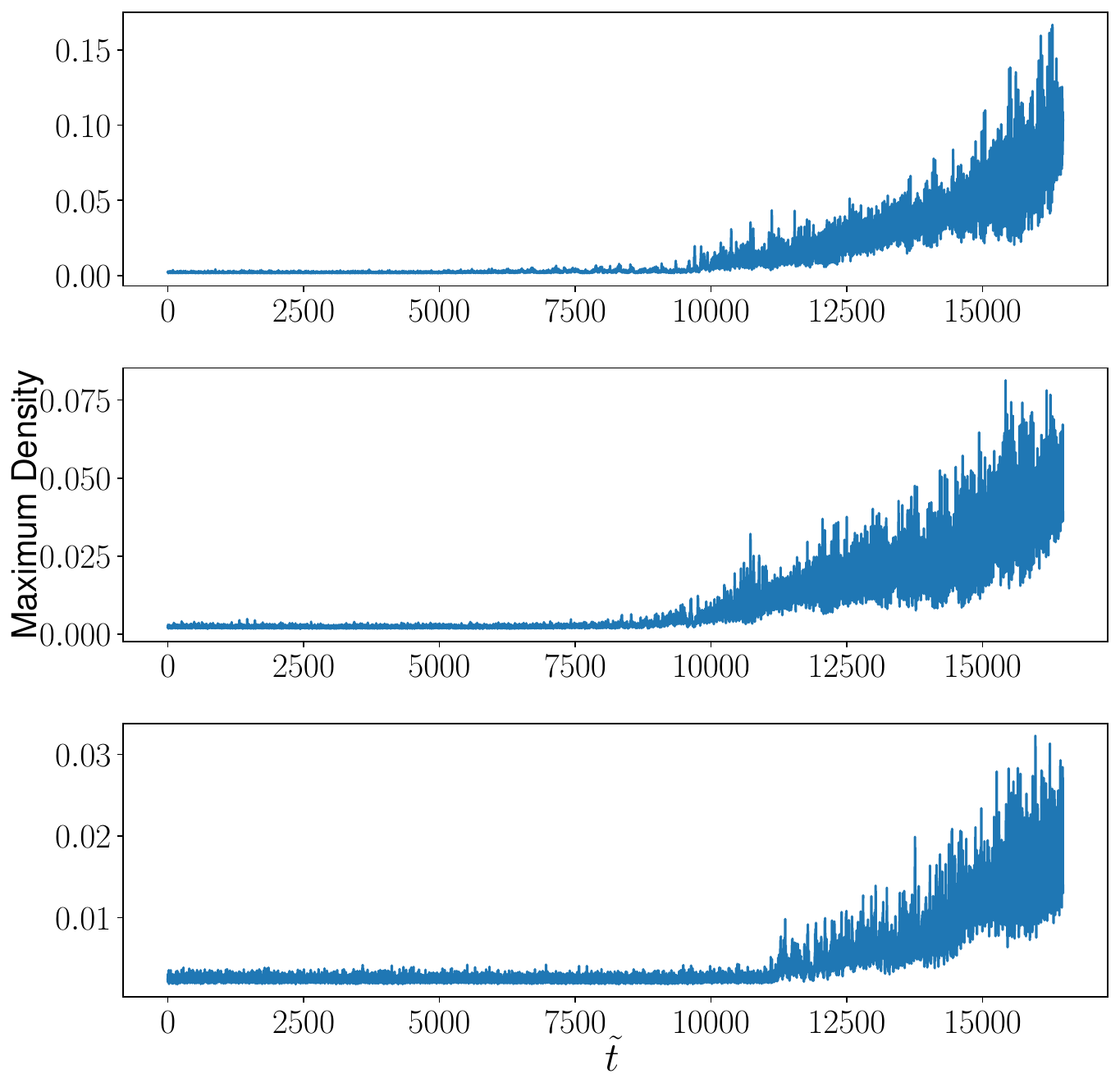}
		\caption{}
		\label{plot6a}
	\end{subfigure}
	\hspace{1cm}
	\begin{subfigure}{6cm}
		\includegraphics[width=6cm,height=8.5cm]{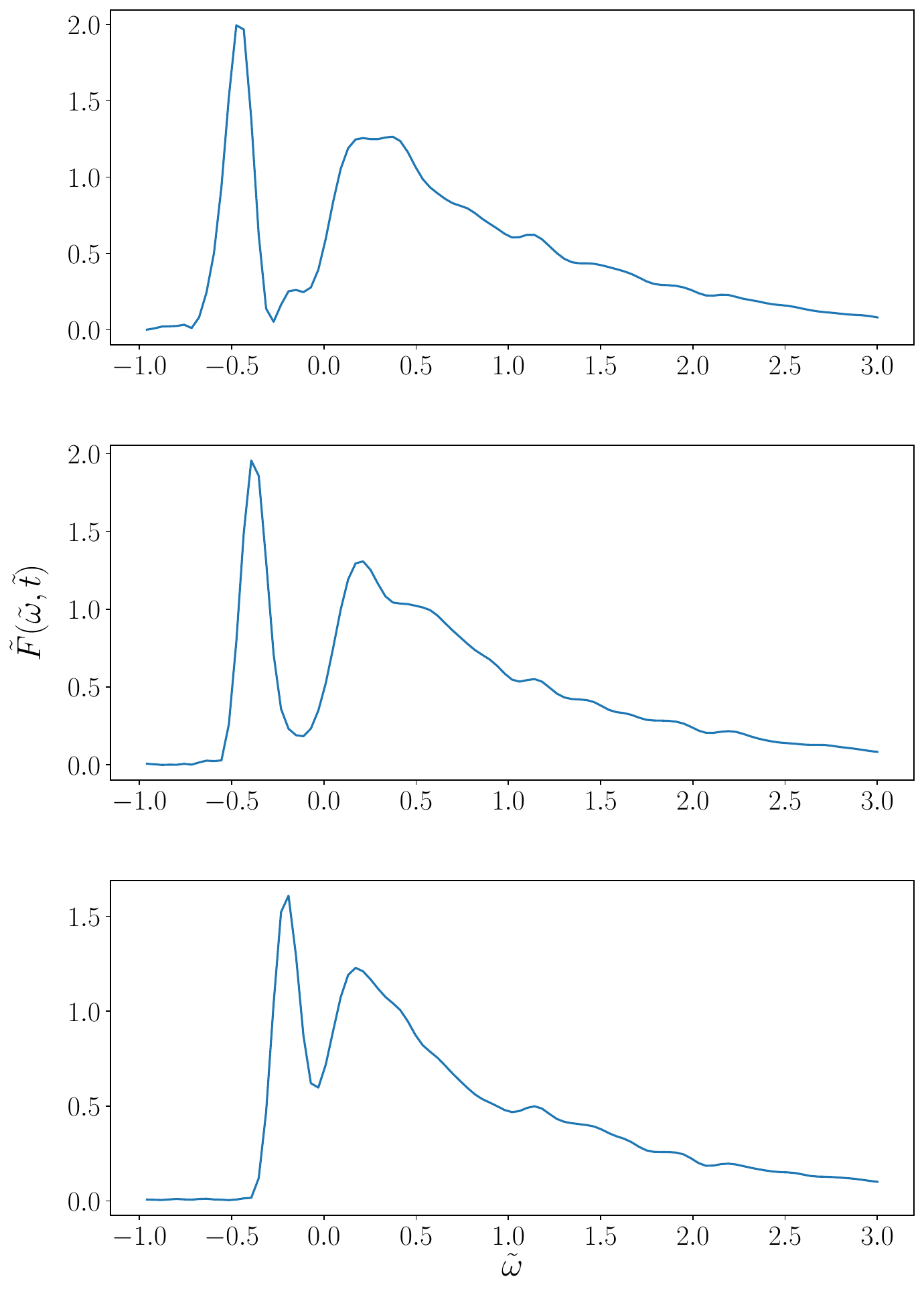}
		\caption{}
		\label{plot6b}
	\end{subfigure}
	\caption{The caption is same as in figure \eqref{plot3}, except here, we have considered attractive self-interaction ($\tilde g=-4.56$) and all plots are obtained at $\tilde t=16000$. }
	\label{plot6}
\end{figure*}

We begin by analyzing the plots of maximum density $\max |\tilde{\psi}|^2$ versus time for different values of $\mathcal{\tilde{L}}_{\rm tot}$, as shown in figure \eqref{plot6a}. All the plots indicate that for a significant duration of time, $\max |\tilde{\psi}|^2$ remains very small and constant. However, after a certain time when condensation occurs, $\max |\tilde{\psi}|^2$ starts to rise monotonically until the convergence test related to the total energy stops the process. The specific time at which $\max |\tilde{\psi}|^2$ begins to increase depends on the value of $\mathcal{\tilde{L}}_{\rm tot}$ present in the initial cloud.

Next, Figure \eqref{plot6b} presents the plots of the spectrum $\tilde{F}(\tilde{t},\tilde{\omega})$ versus frequency $\tilde{\omega}$ at time $\tilde{t}\simeq 16000$ for three different values of $\mathcal{\tilde{L}}_{\rm tot}$. All three plots in Figure \eqref{plot6b} demonstrate the formation of a gravitationally bound state, indicated by the presence of a peak in the negative frequency regime. As discussed earlier, when the interaction is attractive, there exists an upper bound on the coupling constant $\tilde{g}$ given by the condition \eqref{eq12}. This estimation requires determining the value of $\tilde L_B$. In the case of self-interaction, if the Bose star is formed with net angular momentum, it could be unstable with an instability timescale smaller than $\tilde{\tau}_{gr}$.

\begin{figure*}
	\includegraphics[width=14cm,height=4cm,trim={0 0 1cm 0},clip]{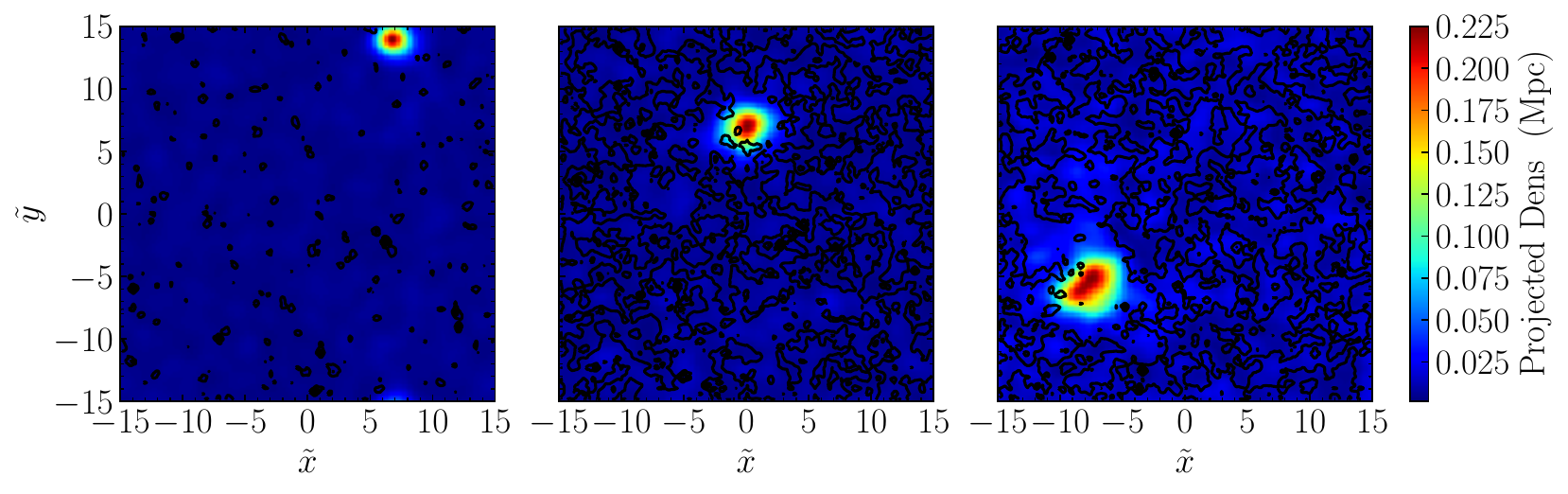}
	\caption{The caption is same as in figure \eqref{plot4}, except here, we have considered attractive self-interaction ($\tilde g=-4.56$) and all plots are obtained at $\tilde t=16000$.}\label{plot7}
\end{figure*}

Figure \eqref{plot7} displays three snapshots of $|\psi|^2$ together with the vorticity magnitude at time $\tilde{t}\sim 16000$ in the $xy$-plane for three different values of $\mathcal{\tilde{L}}_{\rm tot}$. The vorticity magnitude is depicted by the black contours, while the bright red spots indicate the region where the Bose star has formed. The plots reveal that the black contours are concentrated outside the star-forming region. Thus, in this case, similar to the $\tilde{g}=0$ case, we believe that the formed star does not have any intrinsic angular momentum, and our results demonstrate stable stars. These findings align with the stability analysis presented in \cite{Dmitriev}.

\begin{figure*}
	\includegraphics[width=14cm,height=8cm]{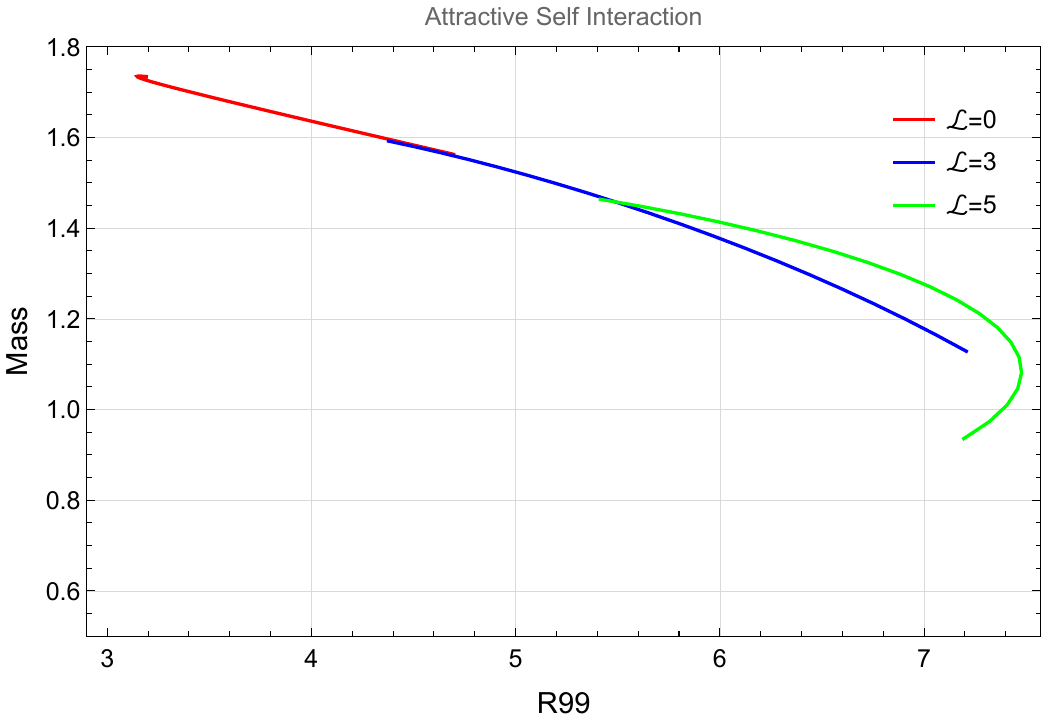}
	\caption{Mass–radius relation for Bose stars with Attractive self-interaction for different values of $\mathcal{\tilde L_{\rm tot}}$}\label{plot8}
\end{figure*}

Figure \eqref{plot8} illustrates $M-R$ diagrams for different values of the total angular momentum of the star. The plot is generated using the digital filter discussed earlier. The red curve correspond to the case when there is no angular momentum in the cloud. The lowest value of mass corresponds to the time when the condensation begins. The curve indicates that as the star mass grows, it becomes more compact. The blue curve exhibits a similar behavior. The green curve corresponds to the case when the total angular momentum is 5. In this case, for the lower branch, the mass of the star increases with the radius, indicating instability. However, for the upper branch, we observe that increasing the mass of the star leads to a more compact configuration.

\subsubsection{Repulsive self-interaction}

Next, we consider the case of repulsive self-interaction. In this case, it is reasonable to expect that the bound state formed in the numerical results is due to gravity. If the coupling constant is increased to values higher than 4.56, bound states can still be formed. However, studying such bound states would require more computational time than allowed by the convergence test implemented to validate the numerical results. Therefore, for now, we restrict our analysis to $\tilde{g}=+4.56$.

\begin{figure*}
	\centering
	\begin{subfigure}{6cm}
		\includegraphics[width=6cm,height=8.5cm]{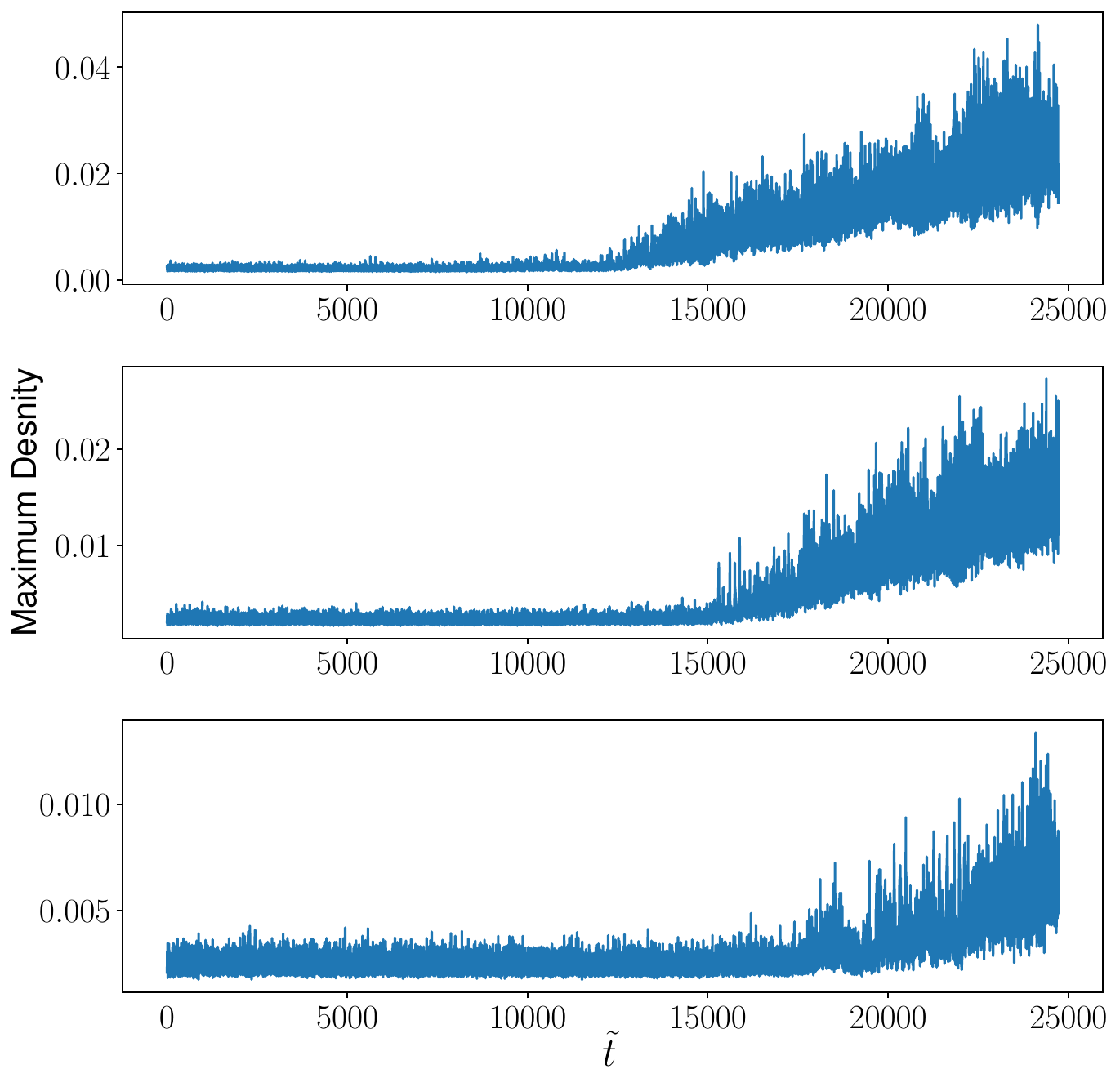}
		\caption{}
		\label{plot9a}
	\end{subfigure}
	\hspace{1cm}
	\begin{subfigure}{6cm}
		\includegraphics[width=6cm,height=8.5cm]{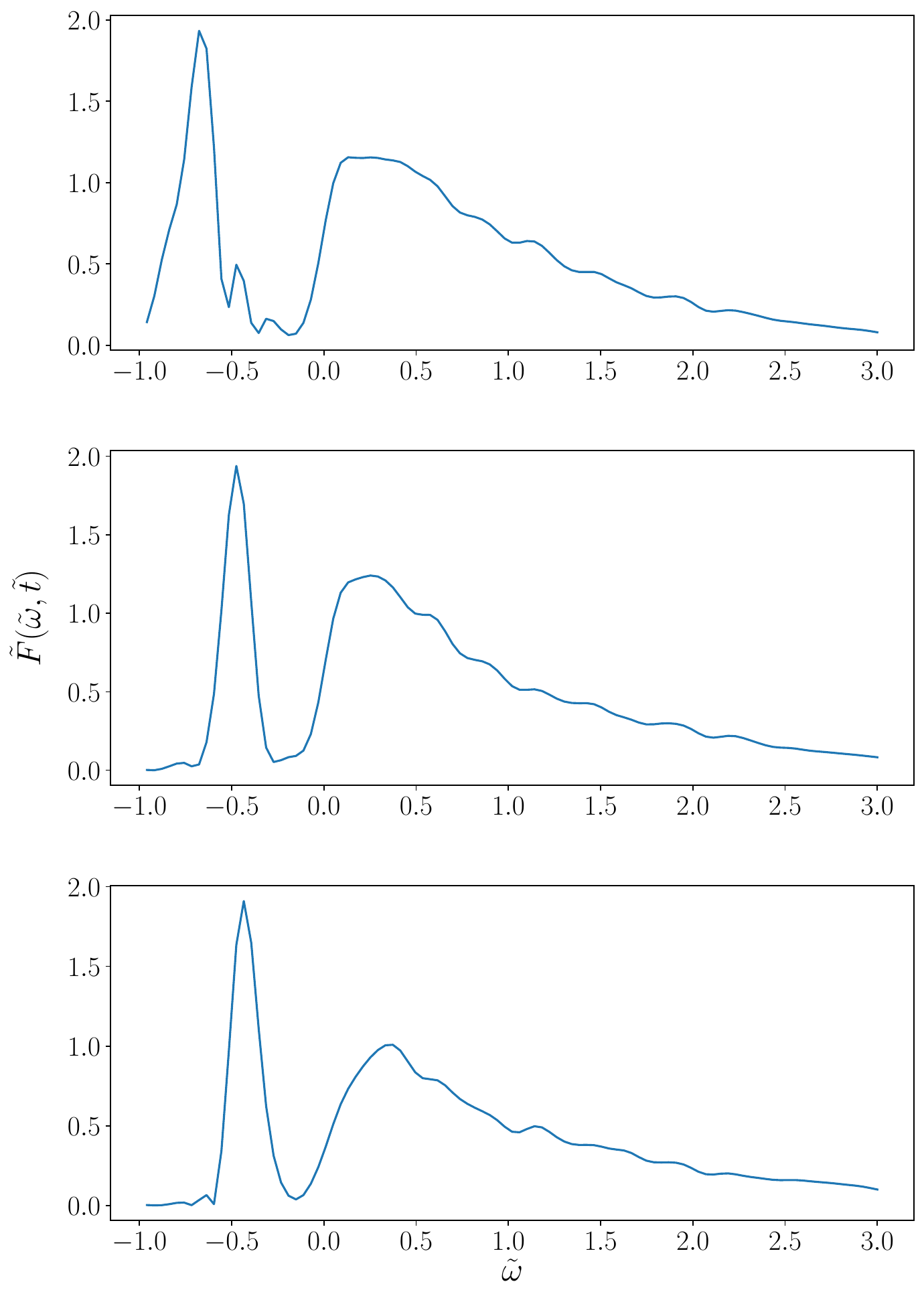}
		\caption{}
		\label{plot9b}
	\end{subfigure}
	\caption{The caption is same as in figure \eqref{plot3}, except here, we have considered repulsive self-interaction ($\tilde g=+4.56$) and all plots are obtained at $\tilde t=25500$. }
	\label{plot9}
\end{figure*}

First, we analyze the plots of the maximum density $\max |\tilde{\psi}|^2$ vs. time, as shown in Figure \eqref{plot9a}. The topmost plot corresponds to the case when the total angular momentum of the collapsing cloud is zero. The middle and bottom plots correspond to the cases with $\mathcal{\tilde{L}}_{\rm tot}=3$ and $\mathcal{\tilde{L}}_{\rm tot}=5$, respectively. All plots show that $\max |\tilde{\psi}|^2$ remains small and constant until time $\tilde{t}\sim\tilde{\tau}_{gr}$. After that, $\max |\tilde{\psi}|^2$ starts to increase with time, until the convergence test limits the program runtime.

Next, Figure \eqref{plot9b} illustrates the plots of the spectrum $\tilde{F}(\tilde{t},\tilde{\omega})$ versus frequency $\tilde{\omega}$ at time $\tilde{t}\simeq, 25500$ for different values of $\mathcal{\tilde{L}}{\rm tot}$. The presence of a peak at negative frequency in $\tilde{F}(\tilde{t},\tilde{\omega})$ indicates the existence of a gravitationally bound state. The stability analysis in \cite{Dmitriev} suggests that an isolated Bose star with angular momentum might be stable. Therefore, it is of interest to investigate what happens to the net angular momentum of the collapsing cloud after the star formation.

\begin{figure*}
	\includegraphics[width=14cm,height=4cm,trim={0 0 0 0.55cm},clip]{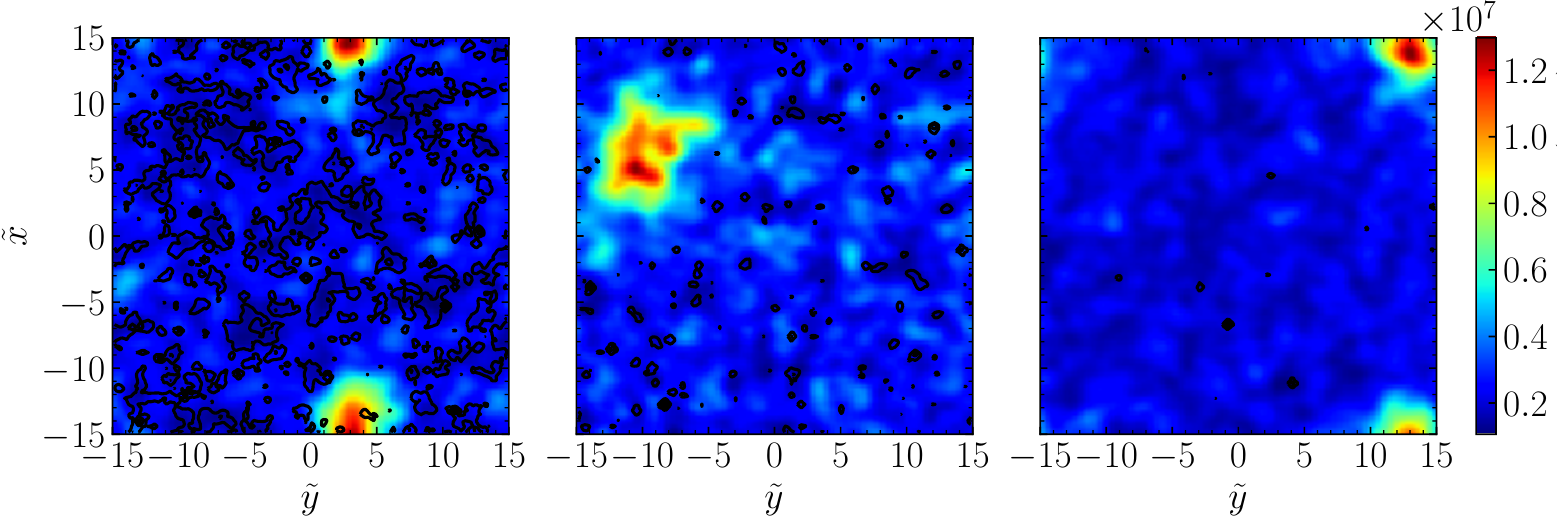}
	\caption{Vorticity magnitude plotted in $\tilde{x}\tilde{y}$ plane at times $\tilde{t} = 22000, 23300\ \textrm{and}\ 27400$ from left to right together with the Density for $\tilde g=+4.56$ (repulsive self-interaction) and $\mathcal{\tilde L_{\rm tot}}\,=\,5.0$. With the passage of time the net vorticity vanishes from the surrounding of the star formation region. The plots does not have any vorticity magnitude outside the star forming region. The absence of vorticity is seen in the other two planes also. This may be indicative of angular momentum entering the star. }\label{plot10}
\end{figure*}

Figure \eqref{plot10} presents three snapshots of density profiles together with the vorticity magnitude when $\mathcal{\tilde{L}}_{\rm tot}=5$. The condensation time for $\mathcal{\tilde{L}}_{\rm tot}=5$ is approximately 18000. The leftmost snapshot, taken at $\tilde{t}\sim22000$, is closer to $\tilde{\tau}_{gr} \sim 18000$. In this plot, the black contours represent the presence of finite vorticity magnitude outside the star-forming region, represented by the red spot. The middle snapshot, taken at time 23300, shows a reduction in the black contours around the star-forming region. This reduction is not limited to the $xy$-plane alone; a significant decrease in vorticity is also observed outside the star-forming region in the other two planes. The rightmost snapshot, taken at time 27400, exhibits a near absence of vorticity magnitude outside the star-forming region. Similarly, vorticity is not observed outside the star-forming region in the other planes. A similar behavior is observed for the $\mathcal{\tilde{L}}_{\rm tot}=3$ case as well, where the vorticity magnitude outside the star-forming region starts decreasing soon after the gravitational condensation.
\begin{figure*}
	\includegraphics[width=14cm,height=8cm]{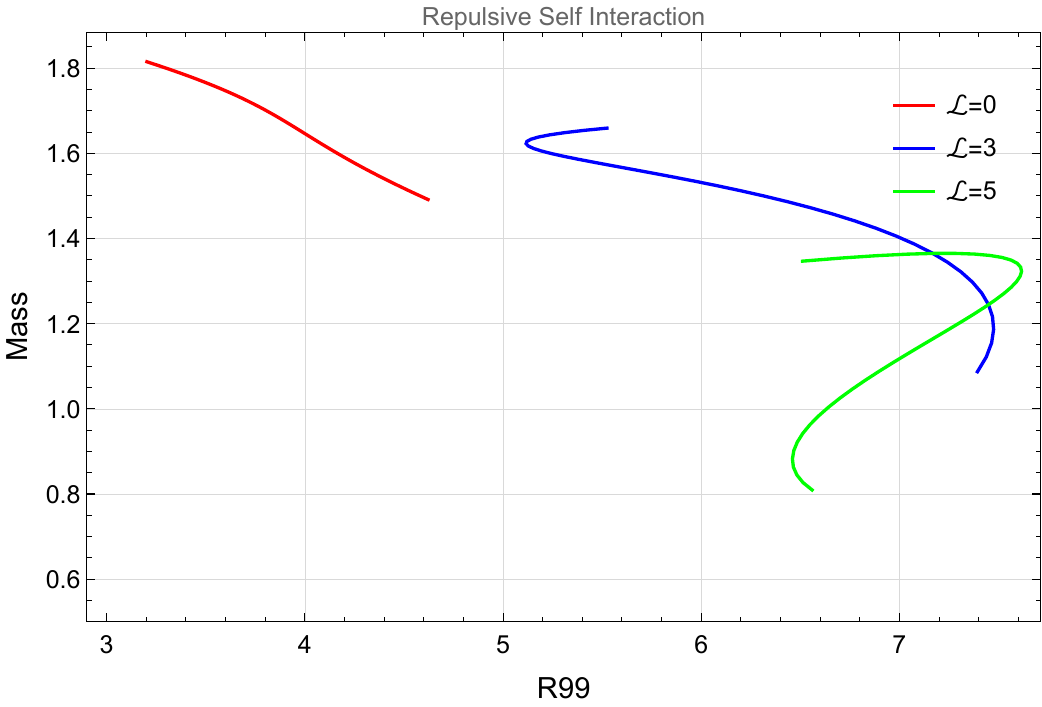}
	\caption{Mass–radius relation for Bose stars with Repulsive self-interaction for different values of $\mathcal{\tilde L_{\rm tot}}$}\label{plot11}
\end{figure*}

Figure \eqref{plot11} shows the $M-R$ diagram when the self-interaction is repulsive for different values of angular momentum. In the absence of angular momentum in the cloud, the red curve clearly indicates that just after the condensation, the star begins with the lowest mass and, as it becomes more massive with the passage of time, it becomes more compact. Interestingly, the vorticity plots after the condensation begins clearly demonstrate that the vorticity vanishes outside the star-forming region. This suggests that the angular momentum might have been transferred to the star. In this case, both the blue and green curves show two branches, where the mass of the star grows with increasing radius, and vice versa. However, there is currently no analytical or numerical work, to the best of our knowledge, that can determine whether a rotating star with repulsive interaction can have a branch of solutions where the mass and radius of the star grow together. These cases are new and require further investigation. However, from the timescales relevant to our simulations, such stars are formed just after the time $\tilde{\tau}_{gr}$ and they survive until the convergence test allows the code to run. In that limited sense, stars represented by the blue and green curves are stable.

\begin{figure}
     \centering
     \begin{subfigure}[b]{\textwidth}
         \centering
         \includegraphics[width=\textwidth]{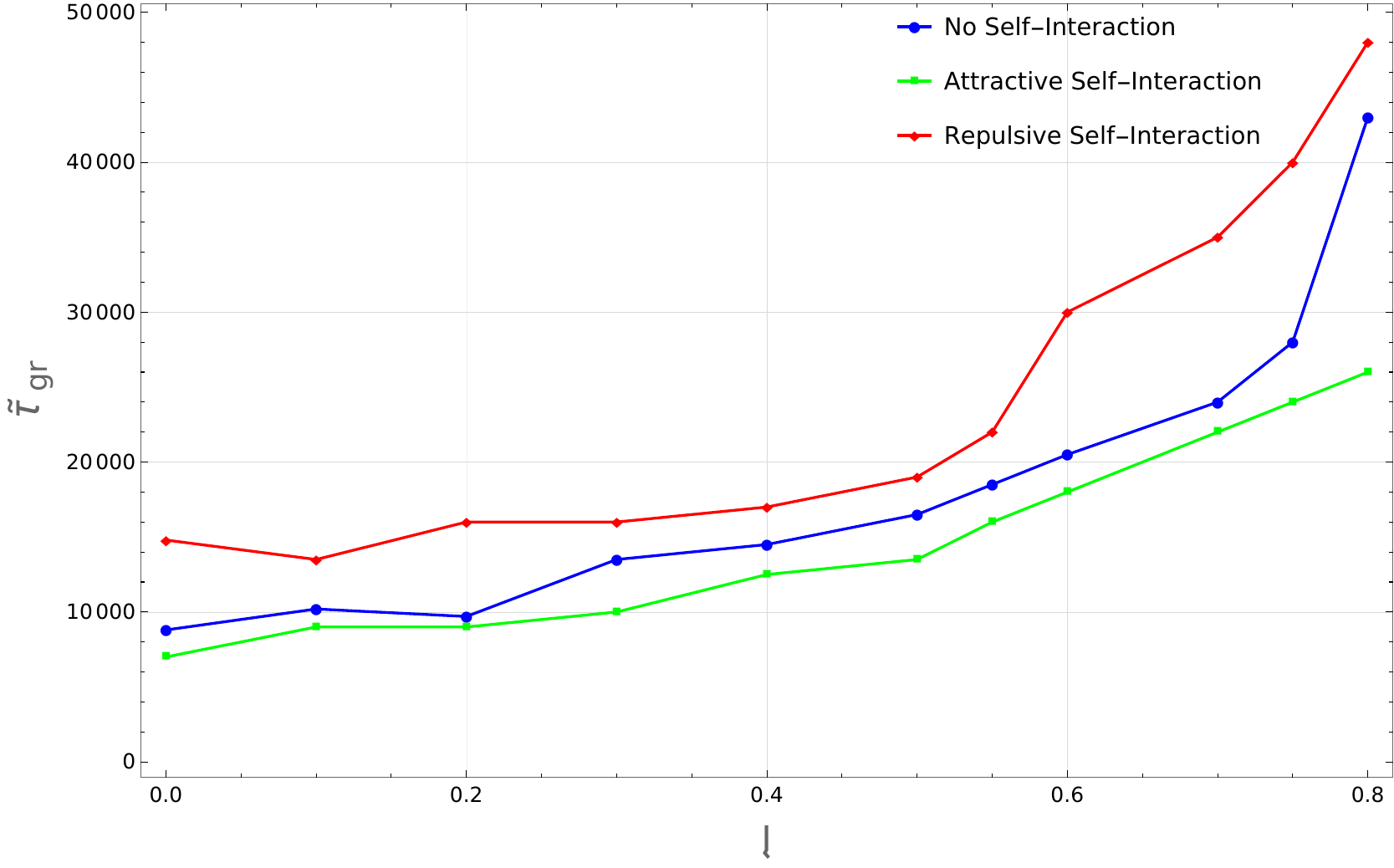}
         \caption{Variation of Condensation Time, ($\tau_{\text{gr}}$), with Angular Momentum per Particle ($l$) for Different Types of Self-Interaction}
         \label{plot12a}
     \end{subfigure}
    % \hfill
     \begin{subfigure}[b]{\textwidth}
         \centering
         \includegraphics[width=\textwidth]{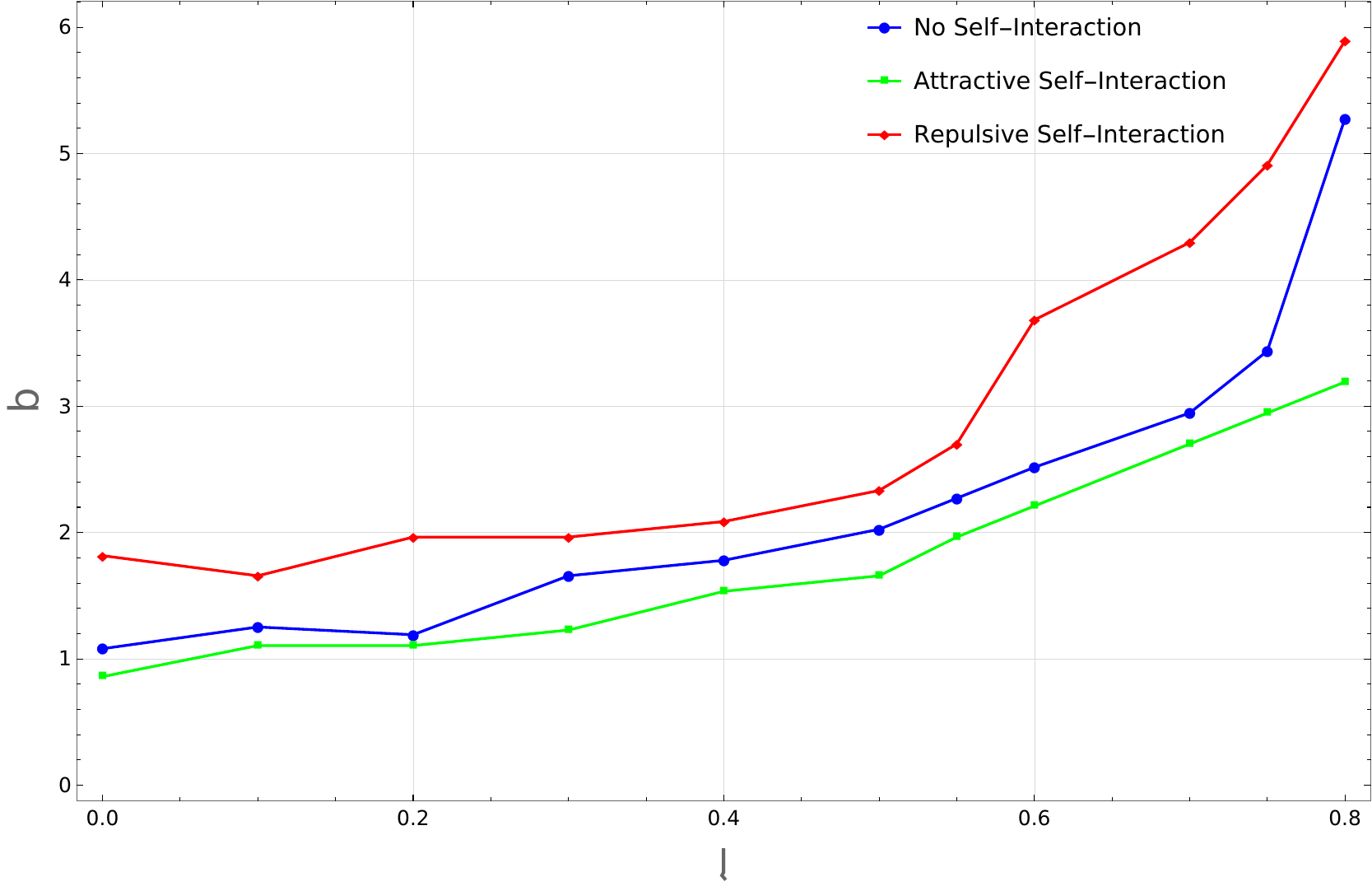}
         \caption{Variation of parameter $b$ with Angular Momentum per Particle ($l$) for Different Types of Self-Interaction}
         \label{plot12b}
     \end{subfigure}
     \hfill
    \caption{}
    \label{plot12}
\end{figure}

In Figure \eqref{plot12a}, we observe the relationship between condensation time ($\tau_{\text{gr}}$) and angular momentum per particle ($l$). Notably, $\tau_{\text{gr}}$ increases in a nonlinear manner with increasing $l$. This trend is consistent across cases involving attractive, repulsive, and no self-interaction. Furthermore, it's evident that, for a specific value of $l$, $\tau_{\text{gr}}$ is shorter when there's an attractive self-interaction compared to the scenario with no self-interaction. Conversely, when there's a repulsive self-interaction, $\tau_{\text{gr}}$ is longer compared to the case with no self-interaction. This underscores the significant influence of both angular momentum in the initial field and the type of self-interaction between particles on the behavior of $\tau_{\text{gr}}$. Figure \eqref{plot12b} provides insights into estimating the parameter $b$ in equation \eqref{eq16} for various values of $l$. Notably, when $l$ is equal to zero and there's no self-interaction, the parameter $b$ is approximately one. However, as we increase the angular momentum ($l$), the parameter $b$ exhibits a nonlinear increase. Additionally, the type of self-interaction also plays a role in influencing the behavior of $b$.

\newpage
\section{Conclusions}\label{sec4}
In conclusion, this work studies the numerical solutions of the nonlinear Schrödinger and Poisson equations, which describe the formation and evolution of a Bose star when the initial cloud has finite angular momentum. The star formation occurs in all the numerical solutions presented here. The results show that once a Bose star is formed in our numerical simulation, it persists until our code violates the energy conservation test.

Since the work is carried out in the kinetic regime, with an initially randomly distributed dark matter density distribution over the entire box, the typical condensation time is on the order of $10^3$ in dimensionless units. The numerical results demonstrate that the presence of angular momentum and self-interaction can significantly affect the star condensation time. Furthermore, the study shows that the presence of angular momentum and self-interaction in the star-forming cloud strongly influences the mass and radius of the Bose star.

 Further we would like to note here that the star formed when the initial cloud does not carry any intrinsic angular momentum performs a random motion during its growth \cite{Levkov}. This allows the formed star to have "orbital" angular momentum. This in turn complicates the important question related with net angular momentum transferred between the rotating initial cloud and the star formed in the numerical simulation. To gain insight into this problem we have studied how vorticity magnitude profile along with the star.  The vorticity magnitude provide a qualitative idea of angular momentum transfer between the initial cloud and the star. More quantitative analysis of angular momentum transfer between the initial cloud and the star shall be done in a future work.
The vorticity magnitude plotted along with the density profile for the cases related with negligible or attractive interaction indicates that the formed star may not possess any intrinsic angular momentum. 
But for the case of repulsive interaction, our study shows the absence of vorticity outside the star-forming region. This implies that the Bose star formed in this case might be having an intrinsic angular momentum transferred from the initial cloud. Thus we believe that our results about the angular momentum transfer are in agreement with the analytical work carried out in Ref. \cite{Dmitriev} in which the stability of a rotating Bose star with different interaction has been analysed.

\section{Acknowledgments}\label{sec5}
We are grateful to Bodo Schwabe, Mateja Gosenca, and Dilip Angom for helpful discussions. And for their help with understanding and modifying the code \texttt{AxioNyx}. We also extend our appreciation to the anonymous referee for their constructive comments that significantly improved the quality of this research paper. All the computations were accomplished on the Vikram-100 and Vikram-1000 HPC cluster at the Physical Research Laboratory, Ahmedabad.

\bibliography{sn-bibliography}% common bib file
\end{document}